\documentclass[prb,twocolumn,amsmath,amssymb,citeautoscript,showkeys,floatfix,superscriptaddress]{revtex4-2}

\usepackage{chemformula} % Formula subscripts using \ch{}
\usepackage[T1]{fontenc} % Use modern font encodings
\usepackage{siunitx}
\usepackage{amssymb}
\usepackage{amsmath}
\usepackage{graphicx}% http://ctan.org/pkg/graphicx
\usepackage{multirow}% http://ctan.org/pkg/multirow
\usepackage{booktabs}% http://ctan.org/pkg/booktabs
\usepackage{footnote}
\usepackage[flushleft]{threeparttable}

\usepackage{hyperref}% add hypertext capabilities
\hypersetup{
    colorlinks,%
    citecolor=blue,%
    linkcolor=blue,%
    urlcolor=blue
}

\newcommand{\orcid}[1]{\href{https://orcid.org/#1}{\includegraphics[width=8pt]{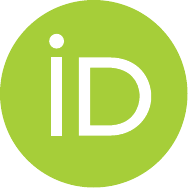}}}

\usepackage{filecontents}

\begin{document}

\title{Defect engineering and Fermi-level tuning in half-Heusler topological semimetals} 
%\title{Cause of excess holes in half-Heusler topological semimetals} 	

\author{Shoaib Khalid\orcid{0000-0003-3806-3827}} 
\affiliation{Department of Material Science and Engineering, University of Delaware, Newark, DE 19716, USA}
\affiliation{Princeton Plasma Physics Laboratory, P.O. Box 451, Princeton, New Jersey 08543, USA}
    
\affiliation{Department of Physics, School of Natural Sciences (SNS), National University of Sciences and Technology (NUST), Islamabad 44000, Pakistan}
\author{Hadass S. Inbar\orcid{0000-0002-8914-1162}}
\affiliation{Materials Department, University of California, Santa Barbara, CA 93106, USA}
\author{Shouvik Chatterjee\orcid{0000-0002-8981-788X}}
\affiliation{Department of Condensed Matter Physics and Materials Science, Tata Institute of Fundamental Research, Homi Bhabha Road, Mumbai 400005, India}
\author{Christopher J. Palmstr\o m\orcid{0000-0003-4884-4512}}
\affiliation{Department of Electrical and Computer Engineering, University of California, Santa Barbara, CA 93106, USA}
\affiliation{Materials Department, University of California, Santa Barbara, CA 93106, USA}
\author{Bharat Medasani\orcid{0000-0002-2073-4162}}
\affiliation{Princeton Plasma Physics Laboratory, P.O. Box 451, Princeton, New Jersey 08543, USA}
\author{Anderson Janotti\orcid{0000-0002-0358-2101}}
\email{janotti@udel.edu}
\affiliation{Department of Material Science and Engineering, University of Delaware, Newark, DE 19716, USA}

%\linenumbers

\begin{abstract}

Three-dimensional topological semimetals host a range of interesting quantum phenomena related to band crossing that give rise to Dirac or Weyl fermions, and can be potentially engineered into novel quantum devices. Harvesting the full potential of these materials will depend on our ability to position the Fermi level near the symmetry-protected band crossings so that their exotic spin and charge transport properties become prominent in the devices. Recent experiments on bulk and thin films of topological half-Heuslers show that the Fermi level is far from the symmetry-protected crossings, leading to strong interference from bulk bands in the observation of topologically protected surface states.
Using density functional theory calculations we explore how intrinsic defects can be used to tune the Fermi level in the two representative half-Heusler topological semimetals PtLuSb and PtLuBi. Our results 
explain recent results of Hall and angle-resolved photoemission measurements. The calculations show that Pt vacancies are the most abundant intrinsic defects in these materials grown under typical growth conditions, and that these defects lead to excess hole densities that place the Fermi level significantly below the expected position in the pristine material. Directions for tuning the Fermi level by tuning chemical potentials are addressed.

\end{abstract}

%\pacs{, }

\maketitle

\section{Introduction} \label{sec:intro}

Half-Heusler (h-H) compounds form a class of ternary intermetallics with diverse electrical and magnetic properties, that includes semiconductors \cite{kandpal2006covalent}, semimetals \cite{katsnelson2008half}, half-metals \cite{tanaka1999spin,hordequin1998half}, and topological semimetals \cite{hirschberger2016chiral,logan2016observation,Liu2016}. Having a structure that can be viewed as zinc-blende with filled tetrahedral interstitial sites and with robust chemical flexibility for occupying the three nonequivalent sites allows for tuning these properties over wide ranges. Recent reports on h-H semimetals with band structures featuring non-trivial topology, such as in PtLuSb \cite{logan2016observation,lin2010half,Shouvik2021} and PtLuBi \cite{Liu2016,Super2013,shekhar2015}, generated great interest in exploring charge and spin transport properties for novel quantum device applications. 
In general, topological semimetals are classified as Dirac, Weyl, or node-line semimetals, characterized by symmetry-protected surface states with the presence of band touching points (Dirac points) or line nodes. Having the Fermi level placed at or sufficiently close to these topological features is key to the observation and utilization of the exotic topological properties in devices. More often than not, these features are either buried deep in the occupied valence bands or too high in energy within the empty conduction bands, far away from the Fermi level. Therefore, finding ways to tune the Fermi level in the desired direction, by adding impurities (doping) or adjusting the concentration of intrinsic defects via growth conditions, minimally perturbing the underlying topological band structure, is crucial for harvesting the full potential of these interesting class of materials.

\begin{figure}
\includegraphics[width=8 cm]{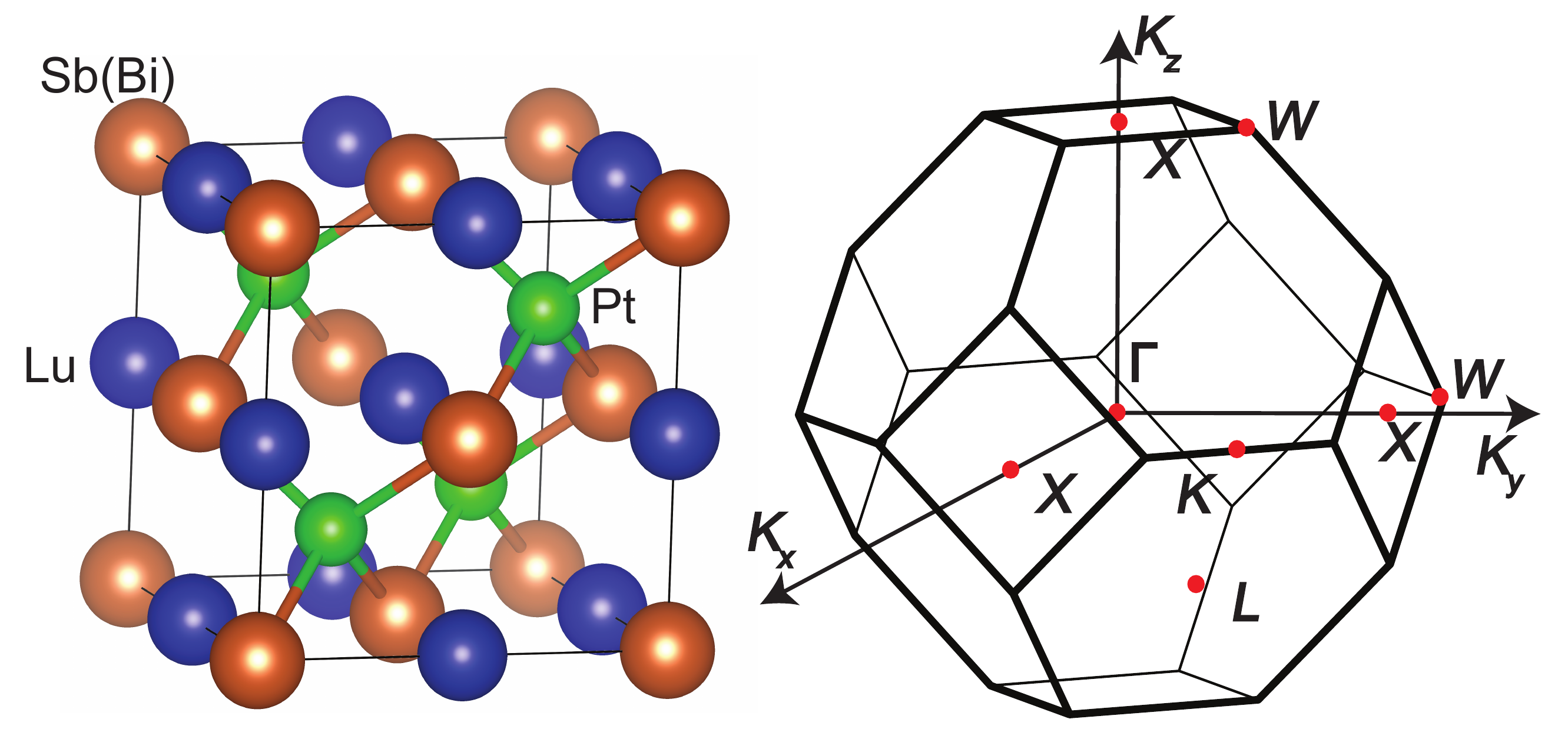}
\caption{Crystal structure 
of half-Heuslers PtLuSb and PtLuBi, viewed as three interpenetrating face centered cubic sub lattices, where the green, blue, and brown denote Pt, Lu and Sb(Bi) atoms, respectively. Pt and Sb(Bi) form a zinc blende sublattice, as indicated by the tetrahedral bonding.  The first Brillouin zone corresponding to the primitive unit cell of the half-Heusler crystal structure, containing one formula unit, is shown on the right.}
\label{fig1}
\end{figure}

\begin{figure*}
\includegraphics[width=18 cm]{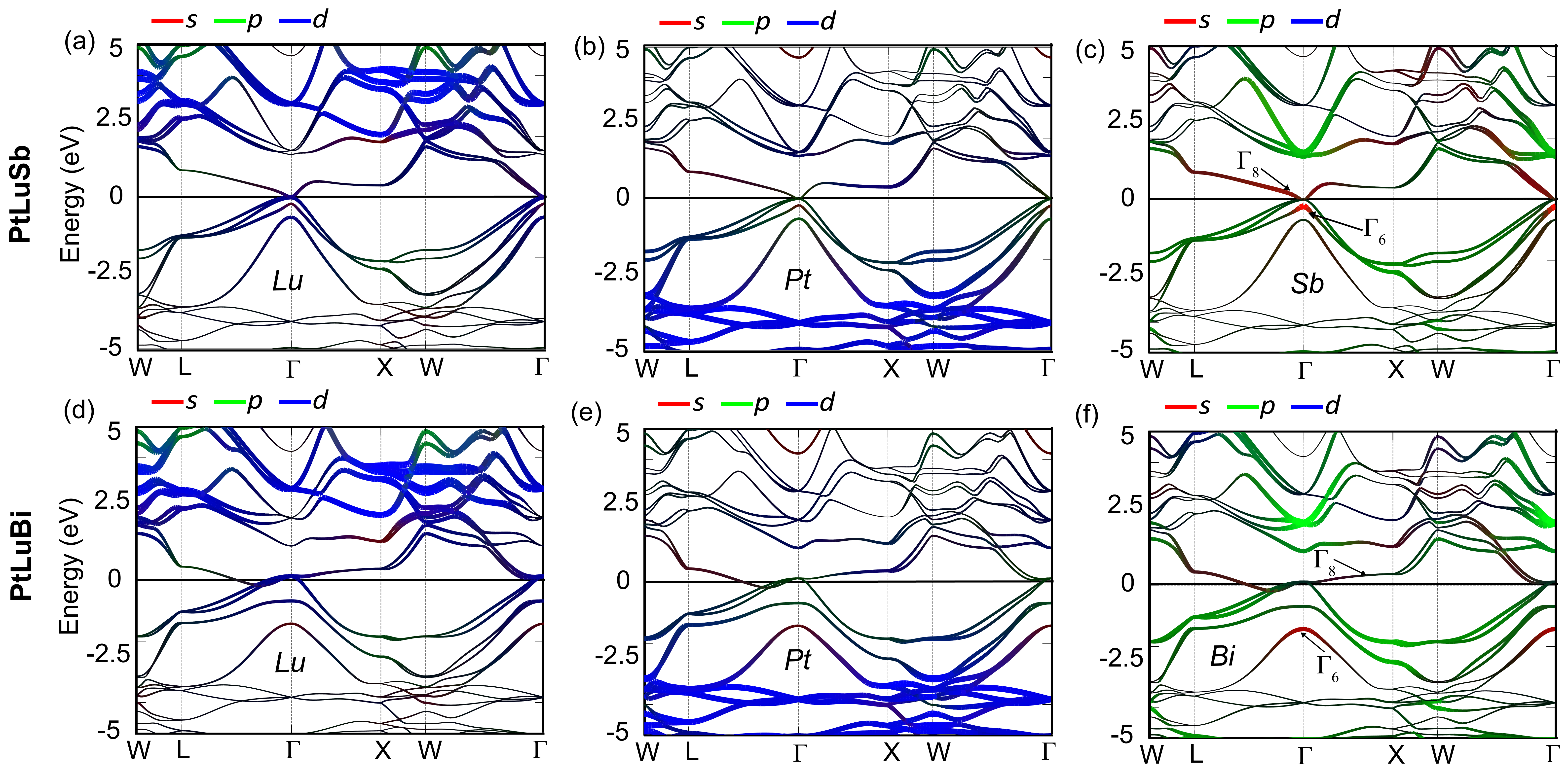}
\centering
\caption{Orbital-resolved electronic band structures of (a-c) PtLuSb and (d-f) PtLuBi. The Fermi level is set to zero, and corresponds to very low carrier densities in the pristine (defect free) material.}
\label{fig2}
\end{figure*}

In the case of PtLuSb and PtLuBi, Hall measurements \cite{logan2016observation,Super2013} show hole conductivity with sizeable hole densities of 3$\times$10$^{20}$ cm$^{-3}$ and 2 - 4$\times$10$^{19}$ cm$^{-3}$, a sign that the Fermi level lies well below what one would expect for the pristine, defect free crystal.  Angle-resolved photoelectron spectroscopy (ARPES) measurements further indicate that the Fermi level is well below the Dirac point of the topological surface states in PtLuSb \cite{logan2016observation}, and well above the Dirac point in PtLuBi \cite{Liu2016}. 
Pronin \textit{et al.}\cite{hutt2018linear} report results of optical conductivity indicating the presence of a triple point in PtGdBi, but this triple point lies well above the Fermi level in the samples due to unintentional $p$-type doping\cite{yang2017}, also seen in the ARPES data \cite{Chang2011}.

The origin of the extra charge carriers in PtLuSb, PtLuBi, and PtGdBi, whether from intrinsic defects or impurities, is still unknown.  Using first-principles calculations based on the density functional theory we investigate the impact of intrinsic defects on the Fermi level position with respect to that expected in perfect crystalline material, searching for an explanation for the observed extra holes and the Fermi level position in these materials. We also lay out design principles to tune the Fermi level by adding specific impurities or changing the atomic chemical potentials during growth or processing to control the concentration  of specific intrinsic defects that  affect the Fermi level position in the desired direction.  We discuss how the presence of Pt vacancies in both PtLuSb and PtLuBi can explain the observed excess hole carrier densities, and address how this effect can be reversed and the Fermi level controlled by adding impurities.

The calculations of intrinsic defects in the topological semimetals PtLuSb and PtLuBi are based on the density functional theory \cite{hohenberg1964inhomogeneous,kohn1965self} within the generalized gradient approximation \cite{Pbesol} and the projector augmented wave (PAW) potentials \cite{blochl1994projector}, as implemented in the VASP code\cite{kresse1993ab,kresse1994ab}. Equilibrium lattice parameters and band structures were calculated using the primitive cell containing three atoms, with a 16$\times$16$\times$16 $\Gamma$-centered mesh of $k$ points for the integration over the Brillouin zone. For the defect calculations, we used a cubic supercell containing 96 atoms which corresponds to a 2$\times$2$\times$2 repetition of the conventional 12-atom cubic unit cell of PtLuSb(Bi), 
%with a total of 32 atoms Lu,Pt and Sb(Bi), 
with $\Gamma$-centered 4$\times$4$\times$4 mesh of special $k$ points. All calculations were performed using a kinetic-energy cutoff of 400 eV for the plane-wave basis set. The 4$f$ electrons of Lu, that form completely occupied bands, were considered as core electrons. Previous calculations of band structures of LuSb and LuBi show that the occupied 4$f$ bands are well below the Fermi level (by more than 8 eV), in agreement with photoelectron spectroscopy measurements in PtLuBi \cite{Liu2016}, and do not affect the structural properties or electronic band structure near the Fermi level \cite{Khalid2020}. 
%Tests with supercells containing 324 and 768 atoms show that formation energies change by less than 0.02 eV. 
The effects of spin-orbit coupling are included in all calculations.

The structure of half-Heusler compounds $ABC$ ($A$= Pt; $B$= Lu; $C$= Sb or Bi) can be visualized as three interpenetrating $fcc$ lattices described by a primitive cell containing three atoms, with atoms $A$, $B$, and $C$ located at (0.5,0.5,0.5)$a$, (0.25,0.25,0.25)$a$ and (0,0,0), respectively, where $a$ is the lattice parameter. Thus, Lu and Sb(Bi) form a rock salt structure while Pt and Sb form a zinc blende structure, as shown in Fig.~\ref{fig1}. 
The calculated lattice parameters for PtLuSb and PtLuBi are 6.443 {\AA} and 6.572 {\AA}, in good agreement with the experimental values of 6.457 {\AA} and 6.578 {\AA} \cite{logan2016observation,Shouvik2021,Super2013}.

The electronic structure and related properties of half-Heusler compounds are closely related to the number of valence electrons \cite{anand2018valence}. These compounds exhibit semiconducting properties similar to that of conventional semiconductor GaAs when the total number of valence electrons per formula unit is equal to 8 or 18 (closed shell), following the so called 8 or 18 electron rule. To understand the topological properties in semimetallic members of this family, represented here by PtLuSb and PtLuBi, the band structure of HgTe can serve as a starting point. HgTe is a well known topological semimetal with a band inversion between Hg-$s$ state ($\Gamma$\textsubscript{6}) and Te-$p$ state ($\Gamma$\textsubscript{8}) at the $\Gamma$ point.  The Fermi level in the intrinsic material is located at the point where conduction band touches the valence band at $\Gamma$. Similar band inversion occurs in PtLuSb and PtLuBi, as displayed in the orbital-resolved electronic band structure shown in Fig.~\ref{fig2}(a-f). For PtLuSb the occupied bands near the Fermi level originate mostly from the Sb atom, with the inversion between $\Gamma$\textsubscript{8} ($p$-character) and $\Gamma$\textsubscript{6} ($s$-character) making it a compensated semimetal with a
topologically non-trivial band structure [Fig.~\ref{fig2}(c)], in agreement with previous studies\cite{lin2010half,logan2016observation,sawai2010topological}. The Lu-5$d$ orbitals mostly contribute to the bands ~2-4 eV above the Fermi level, whereas Pt-5$d$ occupied bands are around ~4 eV below the Fermi level. The calculated band inversion strength (BIS) for PtLuSb is 0.23 eV, in agreement with previous studies\cite{lin2010half,logan2016observation,sawai2010topological}. The band structure of PtLuBi is qualitatively similar, with the $\Gamma$\textsubscript{8} state well above the $\Gamma$\textsubscript{6} state, as shown in Fig.~\ref{fig2}(d-f), also in agreement with the previous studies\cite{lin2010half,sawai2010topological}. The calculated band inversion strength (BIS) for PtLuBi is 1.55 eV, being significantly larger than in PtLuSb due, in large part, to the stronger spin-orbit coupling. 

\begin{figure}
\includegraphics[width=7 cm]{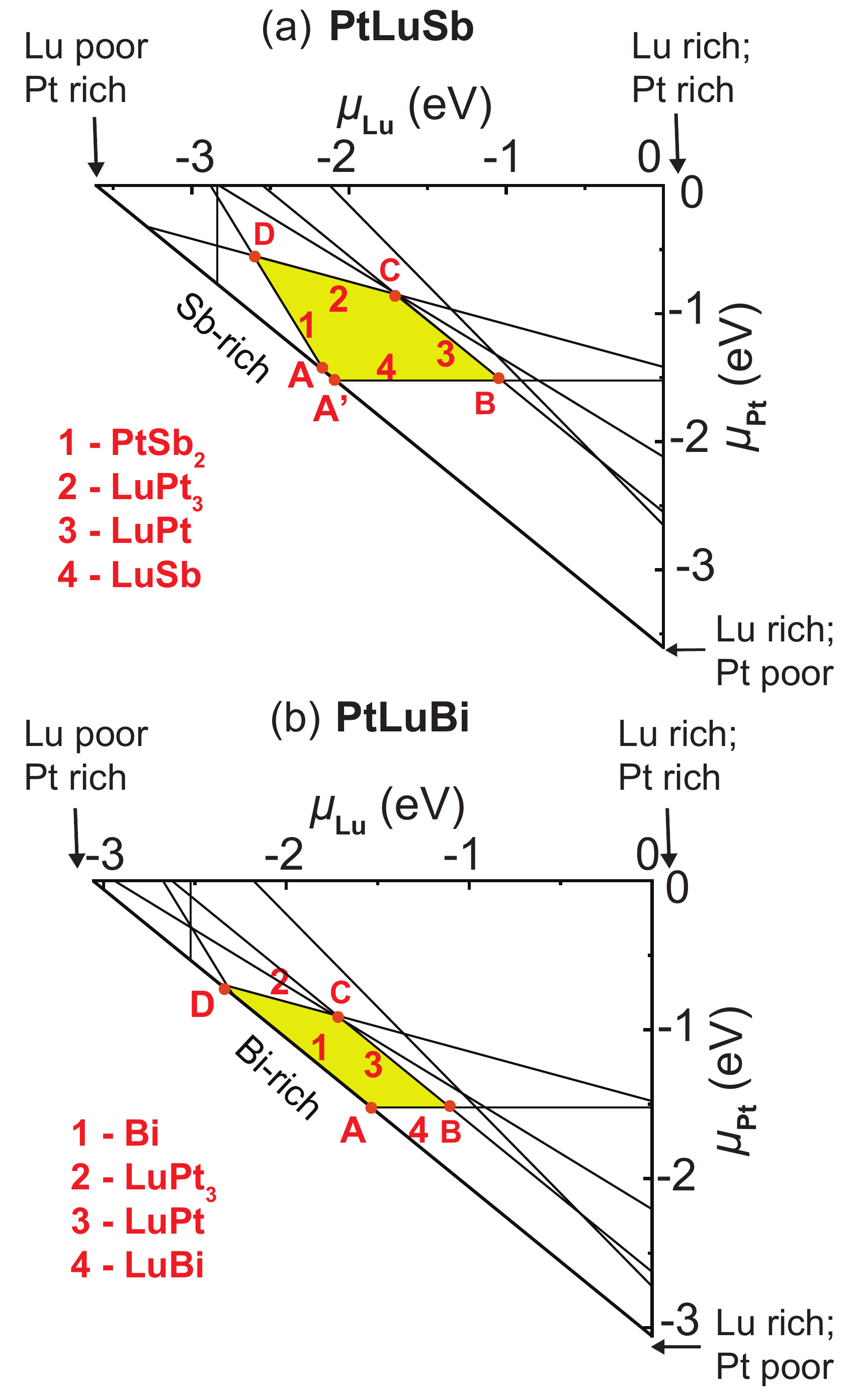}
\caption{The allowed chemical potential region (yellow shading) in $\mu$\textsubscript{Lu} versus $\mu$\textsubscript{Pt} plane along with the competing phases for (a) PtLuSb and (b) PtLuBi.  The point A in both diagrams corresponds to the Sb-rich/Bi-rich limit representing typical growth conditions for bulk and thin films as reported in the literature. }
\label{fig3}
\end{figure}

Defects are modelled by removing, adding, or replacing an atom in a supercell using periodic boundary conditions. 
The results of defect calculations reported here are obtained using a supercell with 96 host atoms. Convergence tests for the lowest energy defects using supercells of 324 and 768 atoms show that formation energies change by less than 0.05 eV when increasing the supercell size, indicating that the 96-atoms supercell is sufficient to simulate an isolated point defect.
The formation energy for a defect $X$ is given by:
\begin{equation}
E^f[X]=E_{tot}[X]-E_{tot}[host]+ n_i \sum\limits_{i} (E_{tot}[X_i] + \mu_i),
\label{equation:formation}
\end{equation}
where $E_{tot}[X]$ is the total energy of the supercell containing the defect $X$, $E_{tot}[host]$ is the total energy of the perfect crystalline host material using the same supercell, $n_i$ is the number of atoms that are removed/added to the supercell to form the defect $X$, and $\mu_i$ is the atomic chemical potential, i.e., the energy of the atomic reservoir for the species added/removed, referenced to the total energy of the respective elemental phases $E_{tot}[X_i]$, and can be related to the experimental growth or processing conditions. The chemical potentials $\mu_i$ are not free parameters, but are bound by the stability condition of the host material (PtLuSb or PtLuBi, in this case) and the need to avoid the formation of possible secondary phases, such as LuSb, LuPt, PtSb$_2$, and LuPt$_3$ in the case of PtLuSb. These conditions are translated into the following relations:
\begin{gather}
\mu_{\rm Lu} + \mu_{\rm Pt} + \mu_{\rm Sb} = \Delta H^f[{\rm PtLuSb}],\\
\mu_{\rm Lu} + \mu_{\rm Sb} < \Delta H^f[{\rm LuSb}],\\
\mu_{\rm Lu} + \mu_{\rm Pt} < \Delta H^f[{\rm LuPt}],\\
\mu_{\rm Pt} + 2\mu_{\rm Sb} < \Delta H^f[{\rm PtSb}_2],\\
\mu_{\rm Lu} + 3\mu_{\rm Pt} < \Delta H^f[{\rm LuPt}_3],
\end{gather}
with $\mu_{\rm Lu}\le 0$, $\mu_{\rm Pt}\le 0$, and $\mu_{\rm Sb} \le 0$ representing the upper limits set by the respective elemental phases. Similar equations are considered for PtLuBi. The range of chemical potentials in the plane $\mu_{\rm Lu}$ vs $\mu_{\rm Pt}$ for PtLuSb and PtLuBi are shown in Fig.~\ref{fig3}, where the relevant regions for the discussion are indicated.  The calculated formation energies of all the native point defects, i.e., vacancies, interstitials, and antisites, on the three sublattices in PtLuSb and PtLuBi are listed in Table~\ref{table:formation1} and Table~\ref{table:formation2} for points A, B, C, and D indicated in the diagrams of Fig.~\ref{fig3}.

\begin{table}
\begin{center}
\caption{Calculated formation energies $E^f$[$X$] of native point defects in PtLuSb at the boundaries of the phase stability diagram in Fig.~\ref{fig3}(a).}
\begin{threeparttable}
\setlength{\tabcolsep}{10pt} % Default value: 6pt
\renewcommand{\arraystretch}{1.5} % Default value: 1
\begin{tabular}{lcccc}
  \toprule\toprule
  \multirow{2}{*}{\raisebox{-\heavyrulewidth}{Defect}} & \multicolumn{4}{c}{Formation energy (eV)} \\
  \cmidrule(lr){2-5}
  & A(Sb-rich) & B & C & D \\
  \midrule
  Sb\textsubscript{Lu} & 1.096 & 3.281 & 2.607 & 1.097  \\
  Sb\textsubscript{Pt} & 4.452 & 5.438 & 6.112 & 5.810  \\
  Pt\textsubscript{Lu} & 4.090 & 5.289 & 3.942 & 2.733  \\
  Pt\textsubscript{Sb} & 3.500 & 2.514 & 1.841 & 2.143  \\
  Lu\textsubscript{Sb} & 4.075 & 1.890 & 2.564 & 4.074  \\
  Lu\textsubscript{Pt} & 6.451 & 5.252 & 6.600 & 7.809  \\
  Sb\textsubscript{i} & 4.501 & 5.558 & 5.558 & 4.953  \\
  Pt\textsubscript{i} & 2.975 & 3.046 & 2.372 & 2.070  \\
  Lu\textsubscript{i} & 5.048 & 3.920 & 4.594 & 5.500  \\
  $V_{\rm Lu}$ & 1.914 & 3.042 & 2.368 & 1.462  \\
  $V_{\rm Pt}$ & 0.635 & 0.564 & 1.238 & 1.540  \\
  $V_{\rm Sb}$ & 4.148 & 3.091 & 3.091 & 3.696  \\
   \bottomrule\bottomrule
\end{tabular}
\label{table:formation1}
\end{threeparttable}
\end{center}
\end{table}

\begin{table}
\begin{center}
\caption{Calculated formation energies $E^f$[$X$] of native point defects in PtLuBi at the boundaries of the phase stability diagram in Fig.~\ref{fig3}(b).}
\begin{threeparttable}
\setlength{\tabcolsep}{9pt} % Default value: 6pt
\renewcommand{\arraystretch}{1.5} % Default value: 1
\begin{tabular}{lcccc}
  \toprule\toprule
  \multirow{2}{*}{\raisebox{-\heavyrulewidth}{Defect}} & \multicolumn{4}{c}{Formation energy (eV)} \\
  \cmidrule(lr){2-5}
  & A(Bi-rich) & B & C & D(Bi-rich) \\
  \midrule
  Bi$_{\rm Lu}$ & 1.325 & 2.188 & 1.530 & 0.568  \\
  Bi\textsubscript{Pt} & 3.867 & 4.298 & 4.916 & 4.718  \\
  Pt\textsubscript{Lu} & 4.226 & 4.658 & 3.386 & 2.619  \\
  Pt\textsubscript{Bi} & 2.966 & 2.534 & 1.917 & 2.115  \\
  Lu\textsubscript{Bi} & 2.446 & 1.583 & 2.240 & 3.202  \\
  Lu\textsubscript{Pt} & 5.515 & 5.083 & 6.358 & 7.122  \\
  Bi\textsubscript{i} & 4.047 & 4.479 & 4.465 & 4.079 \\
  Pt\textsubscript{i} & 2.720 & 2.720 & 2.089 & 1.900  \\
  Lu\textsubscript{i} & 3.627 & 3.195 & 3.840 & 4.415  \\
  $V_{\rm Lu}$ & 1.883 & 2.314 & 1.669 & 1.094  \\
  $V_{\rm Pt}$ & 0.592 & 0.592 & 1.219 & 1.411  \\
  $V_{\rm Bi}$          & 3.726 & 3.295 & 3.308 & 3.695  \\
   \bottomrule\bottomrule
\end{tabular}
\label{table:formation2}
\end{threeparttable}
\end{center}
\end{table}

%The accessible atomic chemical potential region for LuPtSb is relatively larger than for LuPtBi.
In the phase stability diagram of PtLuSb, we find that $V_{\rm Pt}$ has lowest formation energy defect along the line bordering the formation of LuSb (segment 4) in Fig.~\ref{fig3}(a), from Sb-rich (point A') to Lu-rich (point B), whereas Lu$_{\rm Pt}$ has, by far, the highest formation energy. The formation energy of $V_{\rm Pt}$ at point A, also corresponding to the Sb-rich limit, is slightly higher than at point B (0.635 eV vs 0.564 eV).  The formation energies for all the point defects are found to be greater than 1 eV at point C and D. Overall, we predict that $V_{\rm Pt}$ is the most likely defect to form in PtLuSb in the Sb-rich limit, i.e., $\mu_{\rm Lu}=-2.0$ eV, $\mu_{\rm Pt}=-1.5$ eV, and $\mu_{\rm Sb}= 0$ eV (point A). 

In case of PtLuBi, the stability region [Fig.~\ref{fig3}(b)] is relatively smaller yet the results are qualitatively similar to those of PtLuSb. In the region close to point C, all the point defects have formation energies greater than 1 eV. At the point D, all the defects have $E^f$> 1 eV except Bi$_{\rm Lu}$. At points A and B along the LuBi line $V_{\rm Pt}$ is the lowest energy defect with formation energy equal $E^f$=0.592 eV, as listed in Table~\ref{table:formation2}. Moving along the Bi-rich line toward the Lu poor region, we find Bi\textsubscript{$Lu$} to be another point defect with the low formation energy of $E^f$=0.568 eV. So, for chemical potentials near the Bi-rich and Pt-poor lines (near the LuBi phase, segment 4), the lowest energy defect is by far the Pt vacancy, as in PtLuSb.
Experimentally, thin-film growth of PtLuSb by molecular beam epitaxy (MBE) \cite{logan2016observation,Shouvik2021} or bulk growth of PtLuBi \cite{shekhar2015,zhipeng2015} were carried out in Sb/Bi-rich environments, and were most likely to happen near point A in the stability diagrams of Fig.~\ref{fig3}. 

Given the prevalence of the Pt vacancy as the defect with the lowest formation energy in large part of the stability phase diagram of PtLuSb and PtLuBi, and, in particular, near the region where growth/deposition is carried out, we expect it to play major role in determining the electronic characteristics of the reported bulk and thin films. Considering that Pt and Sb compose a zinc blende sublattice within PtLuSb, removing a Pt atom leads to Sb dangling bonds, and according to the electronic band structures in Fig.~\ref{fig2}, we expect Sb dangling bonds to be located near center of the Sb-related 5$p$ bands just below the Fermi level, and lead to excess hole carriers, i.e., a predominant $p$-type behavior, compared to the compensated pristine perfect material where electron and hole concentrations are the same.  Similarly, Pt vacancies in PtLuBi would also lead excess holes. This is consistent with the behavior of Pt vacancies in half-Heusler semiconductors, which were predicted to act as acceptors \cite{yonggang2017natural}. Our results can explain recent ARPES and charge transport measurements in PtLuSb\cite{logan2016observation,Shouvik2021} and PtLuBi\cite{shekhar2015,zhipeng2015}, where excess holes have been observed in thin films and bulk single crystals grown or deposited under Sb- and Bi-rich conditions.

Each Pt vacancy in PtLuSb and PtLuBi is then expected to result in three holes, considering the Sb$^{3-}$ oxidation state in the pristine materials.  We can estimate the Pt vacancy concentration in experiments by assuming that these defects are incorporated during growth (at high temperatures) and they are frozen in as the samples are cooled down for the characterization at room or lower temperatures. Investing the equation for the concentration of Pt vacancies ($c= N_{sites}\exp(-E^{f}/k_{\rm B}T$), where $N_{sites}$ is the density of sites Sb sites, and we can estimate the formation energy of the defect. 

Taking the measured hole concentrations in PtLuSb of 2-3x10$^{20}$ cm$^{-3}$  \cite{Shouvik2021,logan2016observation}, and 2-4x10$^{19}$ cm$^{-3}$ in PtLuBi \cite{shekhar2015,zhipeng2015}, we obtain vacancy formation energies of 0.375 eV for PtLuSb (thin-film grown at T=450$^{\circ}$C\cite{Shouvik2021,logan2016observation}) and 0.562 eV for PtLuBi (bulk grown at T=650$^{\circ}$C\cite{shekhar2015,zhipeng2015}). These values are close to the calculated formation energies near the A points in the phase diagrams of Fig.\ref{fig3}(a) and (b), close to the Sb- or Bi-rich limits. Better agreement is found for the Pt vacancy in PtLuBi, corresponding to bulk growth at higher temperatures, i.e., closer to the thermodynamic equilibrium.

\begin{figure}
\includegraphics[width=8 cm]{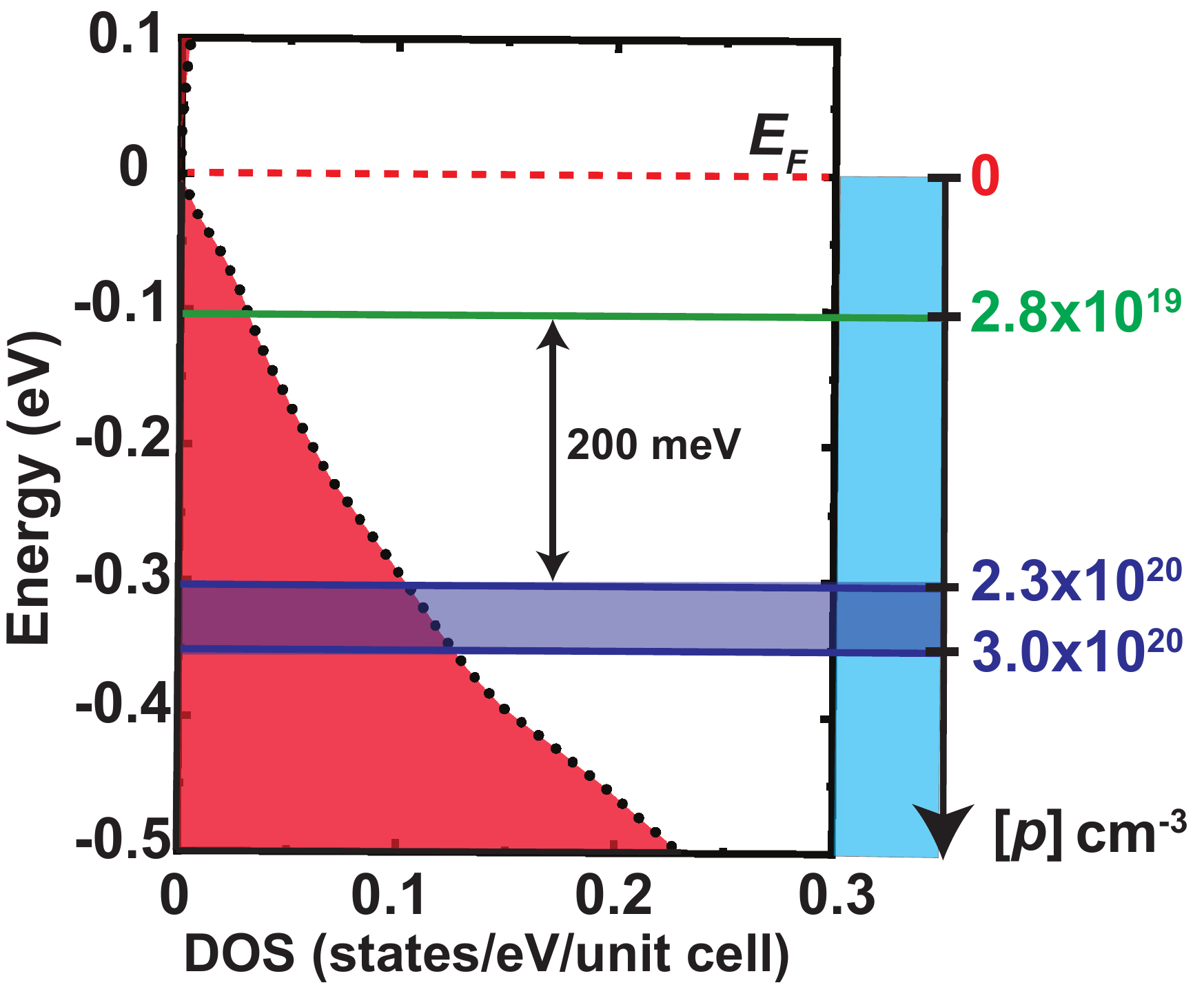}
\caption{(color online) Calculated density of states (DOS) and hole concentration [$p$] as function of Fermi level in PtLuSb, taking the Fermi level $E_F$ in the compensated perfect material as reference. The blue region between the two horizontal lines represents the uncertainty in measured hole concentration. The green line represents the hole carrier concentration after the Fermi level is shifted upwards by 200 meV.}
\label{fig4}
\end{figure}

Such hole concentrations in PtLuSb thin films and PtLuBi bulk single crystals place the Fermi level at significantly lower positions compared to the expected value in the pristine materials, as shown in Fig.~\ref{fig4} for the case of PtLuSb. 
For the measured hole concentrations of 2.3 - 3.0x10$^{20}$ cm$^{-3}$ \cite{Shouvik2021,logan2016observation}, The Fermi level is positioned at 305-360 meV below the expected Fermi-level position in the ideal material.  In order to shift the Fermi level up, towards that of the pristine material and close to the position of the Dirac point \cite{Shouvik2021,logan2016observation}, the Pt vacancy would have to be reduced considerably.  For example, in order to shift the Fermi level upward by 200 meV, the hole density would have to decrease to 2.8x10$^{-19}$ cm$^{-3}$. Considering that these holes come from Pt vacancies and each Pt vacancy contributes with 3 holes, decreasing the hole density is equivalent to a decrease (increase) in the defect density (defect formation energy).
In this case, Pt vacancy formation energy would have to be raised by 132 meV. This can be done by raising $\mu_{\rm Pt}$ by the same amount ($\Delta \mu_{\rm Pt}=132$ meV, i.e., upward from segment 4 in Fig.\ref{fig3}(a)).

In an attempt to translate this number to the knobs in the experimental setup during growth (e.g., Lu/Pt/Sb fluxes in the MBE chamber or sputtering chamber in the case of thin films), we can estimate the change in the Pt partial pressure assuming an ideal gas. We note that this is a simplification of the experimental situation; however, in the case of bulk and thin films, one can imagine that during growth the Pt atoms at the surface are in equilibrium (or close enough to equilibrium) with a flux of Pt atoms, which is idealized here as an ideal gas. Keeping this simplification in mind, we estimate that to raise $\Delta \mu_{\rm Pt}=132$ meV (corresponding to shifting up the Fermi level by 200 meV), an increase in the Pt partial pressure during growth by factor of 8 is necessary.

These results therefore provide an strategy for tuning the Fermi level in these materials. Since Pt vacancy is by far the lowest energy defect in most part of the allowed chemical potential region, increasing the chemical potential $\mu_{\rm Pt}$ is the most direct way of increasing the Pt vacancy formation energy, and therefore, decrease the hole density towards the pristine material.  Another way, is to use aliovalent species that act as donors, filling the Pt vacancies and resulting in extra electrons.  This will also raise the Fermi level, eventually beyond the position in the pristine material, depending on the impurity concentration, and leading to predominant $n$-type conductivity. This effect has been recently reported for Au-doped PtLuSb\cite{Shouvik2021} and can also be explained by the present calculations. Therefore, our results exemplifies how the Fermi level can be tuned and have it placed at the desired positions near the Dirac points where the properties associated with the topological band structures of these materials become more prominent.

In summary, we studied the impact of native point defects on the electronic properties of the topological semimetals PtLuSb and PtLuBi using first-principles calculations. 
We find that Pt vacancy is by far the lowest energy defect in most of allowed chemical potential region in which these compounds are stable, and therefore, the most likely point defect to form during growth of thin films or bulk single crystals. The Pt vacancy leads to Sb dangling bond states that lie below the Fermi level, resulting in acceptor-like states and, thus, excess holes in these otherwise compensated semimetals. The calculated formation energies of the Pt vacancy are consistent with the observed hole concentrations in both PtLuSb and PtLuBi in undoped materials (in which impurity concentrations are negligible). These results also applies to PtGdBi, where the Fermi level was found well below the triple point, and provide a guide to the experiments to refine and pinpoint individual point defects that can affect the transport and magnetoresistance properties in these topologically non-trivial half-Heusler compounds.

%\subsection*{Acknowledgements}
This work was supported by the US Department of Energy Contract no. DE-SC0014388. S.K. and B.M. were supported by the Laboratory Directed Research and Development (LDRD) Program (Grant No. PPPL-132)  at Princeton Plasma Physics Laboratory under  U.S. Department of Energy Prime Contract No. DE-AC02-09CH11466.  The calculations were carried out at the National Energy Research Scientific Computing Center (NERSC), a U.S. Department of Energy Office of Science User Facility operated under Contract no. DE-AC02-05CH11231.
%********************references***********************************
%\bibliographystyle{plain}
\bibliography{LuPtSb-Bi}

\end{document}